\documentclass{iaus}
\usepackage{graphicx,natbib}
\usepackage{bournaud}

\title[Galaxy formation hydrodynamics] 
{Galaxy formation hydrodynamics:\\
From cosmic flows to star-forming clouds}

\author[F. Bournaud]   
{Fr\'ed\'eric Bournaud$^1$}

\affiliation{$^1$ CEA, IRFU, SAp. CEA-Saclay, F-91191 Gif-sur-Yvette, France.\\email: {\tt frederic.bournaud@cea.fr}}

\pubyear{2010}
\volume{270}  
\pagerange{}
\jname{Computational Star Formation}
\editors{J. Alves, B.~G. Elmegreen, J. M. Girart \& V. Trimble, eds.}
\begin{document}

\maketitle

\begin{abstract}
Major progress has been made over the last few years in understanding hydrodynamical processes on cosmological scales, in particular how galaxies get their baryons. There is increasing recognition that a large part of the baryons accrete smoothly onto galaxies, and that internal evolution processes play a major role in shaping galaxies -- mergers are not necessarily the dominant process. However, predictions from the various assembly mechanisms are still in large disagreement with the observed properties of galaxies in the nearby Universe. Small-scale processes have a major impact on the global evolution of galaxies over a Hubble time and the usual sub-grid models account for them in a far too uncertain way. Understanding when, where and at which rate galaxies formed their stars becomes crucial to understand the formation of galaxy populations. I discuss recent improvements and current limitations in ``resolved'' modelling of star formation, aiming at explicitely capturing star-forming instabilities, in cosmological and galaxy-sized simulations. Such models need to develop three-dimensional turbulence in the ISM, which requires parsec-scale resolution at redshift zero.
\end{abstract}

\firstsection 

\section{Cosmological hydrodynamics: the growth of galaxies}

Dark matter-only simulations have long been the main tool of numerical cosmologists. They have drown a picture of galaxy formation in which mass assembly is mostly driven by the gradual merging of small dark matter halos into larger ones: the galaxies, i.e. the baryons, are a relatively passive component that follows this {\em hierarchical } growth of dark halos. 
Cosmological simulations can now directly model the evolution of the baryons, not only in the so-called ``zoom'' simulations towards a single chosen galaxy, but also in large-scale simulations covering hundreds of Mpc, such as the {\em Marenostrum} simulation by Teyssier and collaborators. Such simulations have revealed a richer picture of galaxy assembly: not purely hierarchical, and not driven only by the dark matter structures. Baryons do not all lie within galaxies and halos but also form long filaments along the cosmic web, which gather a large fraction of the gas even at moderate redshift ($z$$\sim$1-2). 

These reservoirs continuously supply gas into galaxies. For small galaxies in low-mass halos, infalling gas is accelerated but the flow remains subsonic in the hot gas that fills the halo (this hot halo gas, coming for instance from stellar winds, is at $T$$\sim$$10^{6-7}$K with $c_s \simeq 100-300$~km/s). The infalling gas thus remains relatively cold ($10^{4-5}$K) and directly fuels the central galactic disk. Around massive galaxies, infalling gas reaches supersonic velocities and is shock-heated to the viral temperature ($10^{6-7}$K) by this {\em virial shock}, which typically stabilizes near the virial radius of the dark halo. These cold accretion vs. hot accretion modes have been proposed by \citet{birnboim03} and detailed in the simulations of \citep{keres05}: the cold mode supplies blue star-forming galaxies, while the hot mode could result in red and dead objects.

Gas infall is not limited to pure cold and hot modes. {\em Cold flows in hot halos} could be a more general mode. Simulations by Kravtsov et al. presented in \cite{dekel06} suggest that cold gas streams at high redshift are dense enough to penetrate the hot gas halo. They directly reach the central galaxy in the form of cold unshocked gas flows collimated in the hot halo. The ubiquity of this mode for massive galaxies was further pointed out in the Marenostrum simulation \citep{ocvirk08}. The number density of such stream-fed galaxies in the simulation matches the observed comoving density of star-forming galaxies at $z \sim 2$ \citep{dekel09}. 
Cold streams in hot halos could be an important growth mode even for relatively low-mass galaxies, as shown by the {\em Nut} simulation (see Powell et al., this meeting). This supports an emerging picture in which Milky Way-like galaxies get most of their mass from cold accretion. Major mergers are relatively unfrequent \citep{genel08}. Even small incoming halos (i.e., minor mergers) account only a small fraction of the total infalling mass, so a substantial part of the infall should be quite smooth \citep{genel09}. The mass brought in as baryonic clumps in dark halos (i.e. major and minor mergers) probably represents, on average, one third of the baryonic supply onto $\sim L*$ galaxies \citep{brooks, keres09}.

\begin{SCfigure}[][t] 
 \includegraphics[width=3.35in, clip]{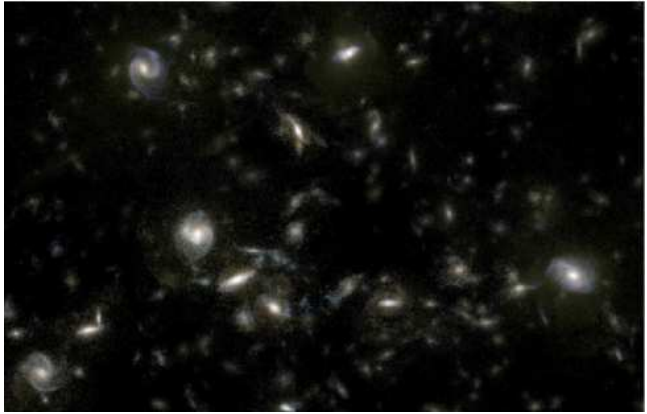} 
\caption{\label{bournaudfig1} {\em Marenostrum galaxies} in a true color synthetic image (courtesy from Teyssier \& Pichon). The main galaxies in this image are around $L*$ and all have bulge/disk ratios around unity. There is a clear lack of bulgeless or disk-dominated galaxies, but also a lack of real ellipticals even in Marenostrum groups and proto-clusters. Sub-grid recipes that would reduce the bulge fraction in all galaxies are not a satisfactory answer to the whole diversity of galaxy populations.} 
\end{SCfigure}

The dominance of mergers or smooth accretion should depend on mass. The galaxy mass function roughly as steep as the dark halo mass function at low masses, and becomes much shallower at high masses \citep{hopkins10}. In the low-mass regime, this implies that galaxy mergers are generally minor ones (having two merging galaxies with similar masses requires two haloes of similar masses, which is unlikely), so cold accretion could dominate. But for very massive galaxies above $L*$, the galaxy masses grow more slowly than the halo masses (because of some debated ``quenching'' mechanism). When two big halos of different masses merge together, their galaxies have more similar masses, and galaxy mergers thus tend to be relatively major ones: giant elliptical galaxies likely mostly assembled through violent mergers.

The formation of dwarf galaxies with stellar masses around $10^9$~M$_{\odot}$, largely driven by cold gas accretion and feedback processes, can be relatively well understood \citep[][and Brook this meeting]{governato}. Giant elliptical galaxies are also relatively well understood as resulting from numerous early mergers and several possible quenching mechanisms (\citealt{naab07}, \citealt{johansson09}, \citealt{martig09}). Things are much more complicated for galaxies in the $L*$ regime, where the combination of cold accretion, mergers, feedback processes and internal evolution is richer. This regime encompasses a large diversity of galaxy types, from bulgeless rotating disks to slowly-rotating ellipticals. While environmental effects play an important role, they cannot explain the bulk diversity of galaxies: there are field ellipticals, and galaxies with very low bulge fractions in groups as well (such as M33 in the Local group). Explaining the diversity of $L*$ galaxies remains a major challenge, the toughest part likely being to explain the survival of bulgeless, pure disk galaxies. For instance, all Marenostrum galaxies at $\sim$$L*$ are bulgy galaxies, say, S0 or Sa (Fig.~1). It is known that all simulations of this type lack bulgeless galaxies, but it is interesting to note that they lack elliptical galaxies at the same time.

\section{Milky Way progenitors in the early Universe}
Since the formation of today's $\sim L*$ galaxies cannot be directly understood from cosmological simulations, let us examine the properties of their progenitors at redshift $z\sim 2$, when they contained only half of their present mass, focusing on normally star-forming galaxies rather than strong starbursts:
Optical surveys can decently resolve the morphology of $L*$ galaxies at $z~2$. An extensive study was performed by Elmegreen, Elmegreen and collaborators (2004--2009) in the Hubble Ultra Deep Field (UDF). Most star-forming galaxies around $L*$ at $z>1$ are very clumpy, with giant  clumps that do not resemble the star-forming clouds or GMCs of nearby spiral galaxies (Fig.~2). These high-z galaxies are most often ``clumpy disks'' rather than multiple mergers. Giant clump masses can reach $\sim$$10^{8-9}$~M$_{\odot}$.
Spectroscopic surveys study the resolved kinematics of ionized gas \citep[e.g.][]{FS09,epinat09}. They have confirmed that the majority of normally star-forming objects in the young Universe are large rotating disks, with giant H$\alpha$ clumps, and suggest strong ISM turbulence with 1D dispersions of a few tens of km/s.
Molecular gas observations of normally star-forming galaxies at $z>1$ \citep{daddi09, tacconi10} have revealed high gas fractions, typically $\sim$50\% of the baryons at $z=2$, with most of the gas most in dense molecular form. Relatively low excitations and Milky Way-like conversion factors are supported by dynamical arguments \citep{daddi09}  and observations of CO SEDs \citep{dannerbauer09}.

\begin{SCfigure}[][t] 
 \includegraphics[width=3.1in, clip]{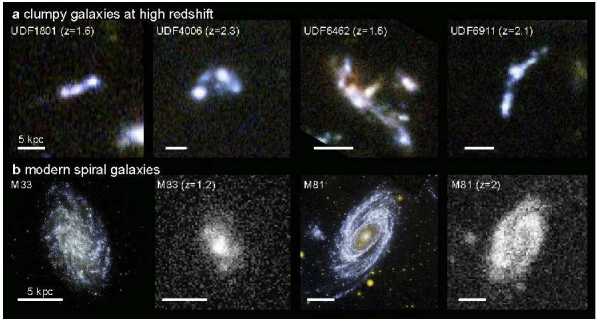} 
\caption[]{\label{bournaudfig2} {\em Star-forming galaxies at $z$$\sim$2} are dominated by kpc-sized clumps of $10^{7-9}$~M$_\odot$. This is not an artifact from bandshifting (optical imaging traces the UV emission at $z=2$) or resolution. Modern disk galaxies in the UV have only much smaller star forming regions. They do not show highly contrasted giant clumps but exponential disks and spiral arms instead, even when artificially redshifted (courtesy from Debra Elmegreen). }
\end{SCfigure}

Gas-rich, turbulent, clumpy disks thus appear to represent the bulk of normally star-forming galaxies at $z>1$. The giant clumps likely result from gravitational (Jeans) instability in gas-rich disks with a Toomre parameter $Q$ below or close to unity, owing to low spheroid fractions, and strong turbulence maintains a high Jeans mass (typical clump mass) \citep{BE09}.

\section{Internal evolution in primordial high-redshift galaxies}

Models of internal dynamics of clumpy disk galaxies were studied in \citet{BEE07}, among others. These clumps are massive enough to undergo tidal torquing and friction and migrate inwards in a few disk rotation periods. They can form a bulge with properties typical for classical bulges (at least in massive galaxies). The clumps loose a substantial fraction of their stars via tidal effects, and maybe some gas through feedback processes. The released material can form an exponential disk. The pre-existing stars can be stirred in a very thick stellar disk, which can be hot enough to remain thick down to $z$$=$0 even if mass infall continues to fuel the thin disk \citep{BEM09}.

\begin{SCfigure}[][t] 
 \includegraphics[width=2.2in, clip]{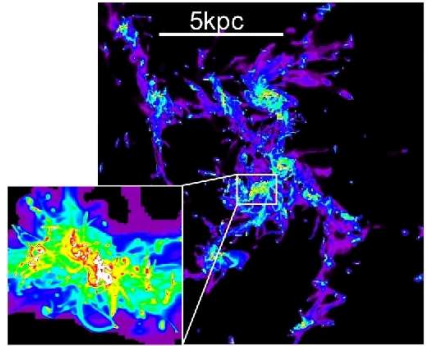} 
\caption[ ]{\label{bournaudfig3} Gas surface density in an AMR model of a gas-rich high-redshift disk with 2~pc resolution, showing a distribution dominated by a few ``giant clumps''. A zoom on a giant clump shows sub-clumps and shocks. This giant complex is overall supported by fragmentation and internal turbulence, and some internal rotation. This prevents the whole complex from reaching very high average gas densities and high star formation efficiencies, which could leave it long-lived against stellar feedback.}
\end{SCfigure}

Rapid internal evolution with bulge formation by giant clump instabilities in high-redshift disks has now been reproduced with various codes and in cosmological simulations \citep{ATM09,CDB10}. A major unknown is the survival of the giant clumps to star formation feedback, in particular radiative pressure from young stars \citep{murray}. \citet{KD10} have shown that giant clumps will survive if they do not collapse into small superdense objects that would form stars too rapidly and undergo too strong feedback. This seems to be the case in observations but measurements at the scale of clumps remain uncertain. This is also the case in AMR disk simulations (Fig.~3) where the giant star-forming complexes form many sub-clumps and remain supported by internal turbulence and rotation instead of collapsing. \citep{burkert09, elm-burkert10} further suggest that the strong turbulence in high redshift galaxies is mostly driven by the gravitational instabilities, not by feedback processes. \citep{burkert08} has shown simulations where the main source of turbulence was a strong disruptive feedback, but the result was not to disrupt the giant clumps after they formed, but to completely inhibit their formation, which is not consistent with observations. There are probably observed signatures of feedback, but they may leave the largest and densest clumps relatively unaffected \citep{lehnert}.

\section{Galaxy mergers and the disk survival issue}
The survival of disk galaxies remains a major issue: even if a large part of the mass comes from smooth flows, galaxies should undergo a significant number of mergers. The survival of massive disk galaxies with low bulge fractions remains misunderstood. Even if giant clumps and subsequent bulge formation at high redshift did not affect galaxies much, merger-induced bulge formation alone seems inconsistent with the observed fractions of bulges \citep{weinzirl}.

There have been claims that disks can survive or re-form in high-redshift mergers, because these mergers involve very high gas fractions \citep[e.g.,][]{robertson06}. Some results were even called ``spiral'' galaxies \citep{SH05} because a large rotating gas disk was found, but more than 50\% of the baryons were actually in a stellar bulge or spheroid after the merger: this is not typical for a ``spiral'' galaxy. A more general issue is that mergers have never been studied in realistic high-redshift conditions, because the employed techniques have always modeled a warm, stable and smooth gas. The real ISM, especially in high-redshift galaxies, is highly unstable, and supported by supersonic turbulence in cold gas. Real mergers at $z$$=$2 with high gas fractions may thus be fundamentally different and much more dissipative than existing models. Disk survival is uneasy with a turbulent, heterogeneous ISM \citep{hzmergers}.

\section{ISM physics in galaxy formation}

Star-forming regions are not just substructures that emit most of the visible light. They have major consequences on disks and bulges at high redshift, and a significant impact in mergers, too \citep{bois, teyssier10}. The crucial role of the heterogeneous, cloudy and turbulent ISM and clustered star formation in global galaxy properties has also been pointed out in recent cosmological simulations. In low-mass dwarf galaxies, this can preserve bulgeless galaxies (Brook, this meeting). More massive galaxies in the $L*$ regime still have too massive bulges and spheroids in cosmological models, but recent work has shown that the physics of ISM turbulence, clustered star formation, and stellar evolution may strongly reduce the predicted bulge fractions \citep[e.g.,][]{semelin,piontek, MB10}. Nevertheless, as pointed out by \citet{ATM10}, too many free parameters in sub-grid models prevent robust predictions, and explicitly resolving the main star forming regions is needed.

\section{Resolved star formation in cosmological and galaxy-sized models}

Galaxy formation models need to directly resolve the largest scales of star formation: a few giant clumps in high-redshift galaxies, and a few tens of GMCs that drive the bulk of the star formation in the Milky Way. In theory, the formation of GMCs is probably initiated by gravitational instabilities in turbulent gas \citep{E02, kmk}, which should be relatively simple to model. However, ``resolving'' star formation is not achieved so easily in galaxy formation models.

In cosmological simulations, none of the recent and supposedly high-resolution models resolves the main star formation regions at $z=0$. Models by Agertz et al. or Governato et al. show an heterogeneous ISM, which mainly results from supernovae feedback creating empty bubbles and higher-density regions. But high densities of thousands of atoms/cm$^3$ are not reached and cloud-forming instabilities are absent. Things are a bit easier at redshift $z>1$ with giant clumps of star formation. They have been directly captured in cosmological simulations \citep{ATM09,CDB10}. They have not been ``resolved'', though: these giant clumps survive feedback because they keep a relatively low density and hence a relatively low star formation efficiency. Maybe this is how a real clump would be, but the resolution limit may induce a severe bias. The smallest cell size in Ceverino et al. is 70~pc, and the thermal Jeans length has to be kept larger than a few cells, the total (thermal+turbulent) Jeans length being necessarily even larger, say, around 500~pc. A star-forming clump could not become much more compact, extremely dense, very efficiently star-forming and rapidly disrupted by feedback. It cannot fragment into sub-clumps and become supported by internal turbulence, but remains largely ``supported by the resolution limit'', instead. Only isolated disk simulations directly resolving the internal fragmentation of such giant clumps is directly resolved and their internal turbulent support, so that their internal density distribution and resulting efficiency in a given star formation scheme can be directly resolved (see Fig.~3).

Resolving star-forming clouds at redshift zero is harder, even in idealized isolated galaxy simulations with box sizes of a few tens of kpc. Recent works have studied the formation of GMCs via cooling and gravitational instabilities in a galactic disk \citep[e.g.][]{li,agertz-33,tasker-tan}. They have highlighted the role of gravity and turbulence in driving GMC formation. However, these simulation do not ``resolve'' the GMC properties: for instance the typical radius of a GMC in the Tasker \& Tan model is around 20~pc, for an AMR cell size of 7~pc. The outer parts of these GMCs are supported by rotation around the central core with a mini-spiral shape (unlike real GMCs) and the internal regions are supported by the resolution limit (i.e., by the thermal pressure floor that maintains the thermal Jeans length larger than a few cells). While the mass of these GMCs could be realistic, their properties such as the gas density distribution or local free-fall time cannot be trusted. Then, assuming a given star formation recipe on small scales (Schmidt law or any other), the resulting star formation activity in each cloud cannot be robustly predicted.

Nevertheless all these simulations have nicely shown that GMC formation is probably mostly driven by ISM turbulence and gravity. We recently went one step further and demonstrated the properties of gas clouds can be directly resolved in simulations of whole galaxies, including their size, density distribution, and internal turbulence (\citealt{bournaud10} and Fig.~4). The average Jeans scale length sets the disk scale height. Starting from this most unstable scale, a 3D turbulence cascade takes place. The turbulence cascade forms denser gas regions, some of which become gravitationally unstable and collapse into dense clouds, with local densities reaching $10^{5-6}$~cm$^{-3}$. This occurs mainly in regions already compactified by stellar arms or bars, which traces the coupled instability of gas and stars in a bi-component disk (see also \citealt{yang07}). Owing to a very high spatial and mass resolution (0.8~pc and $5\times 10^3$~M$_{\odot}$), the model shows GMCs that stopped collapsing well before the resolution limit, and inside which dense sub-clouds, bubbles, and numerous structures with chaotic motions are directly resolved. A single GMC is shown on Fig.~4: its PDF is log-normal with a substantial excess of high-density gas, consistently with observations of nearby GMCs (\citealt{lombardi}). Star formation in the model takes place above 5000~cm$^{-3}$, which is about the high-density tail of the GMC PDF. When collapsing, this GMC evacuated most of its angular momentum by fragmenting and expulsing some substructures, and at the observed instant is dominated by internal turbulent motions.

\begin{figure}
\begin{center}
 \includegraphics[width=1.1in]{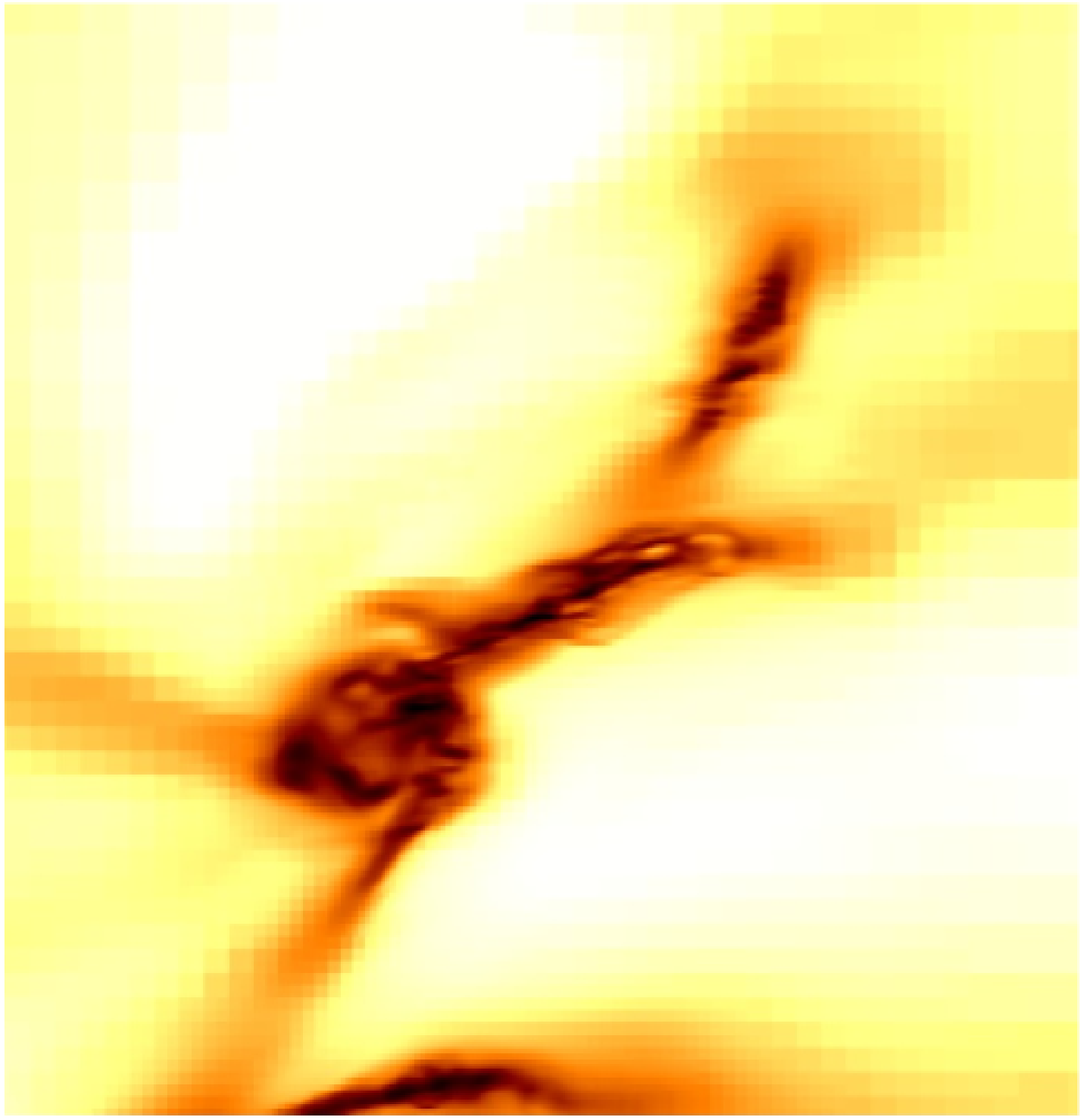} 
 \includegraphics[width=1.7in]{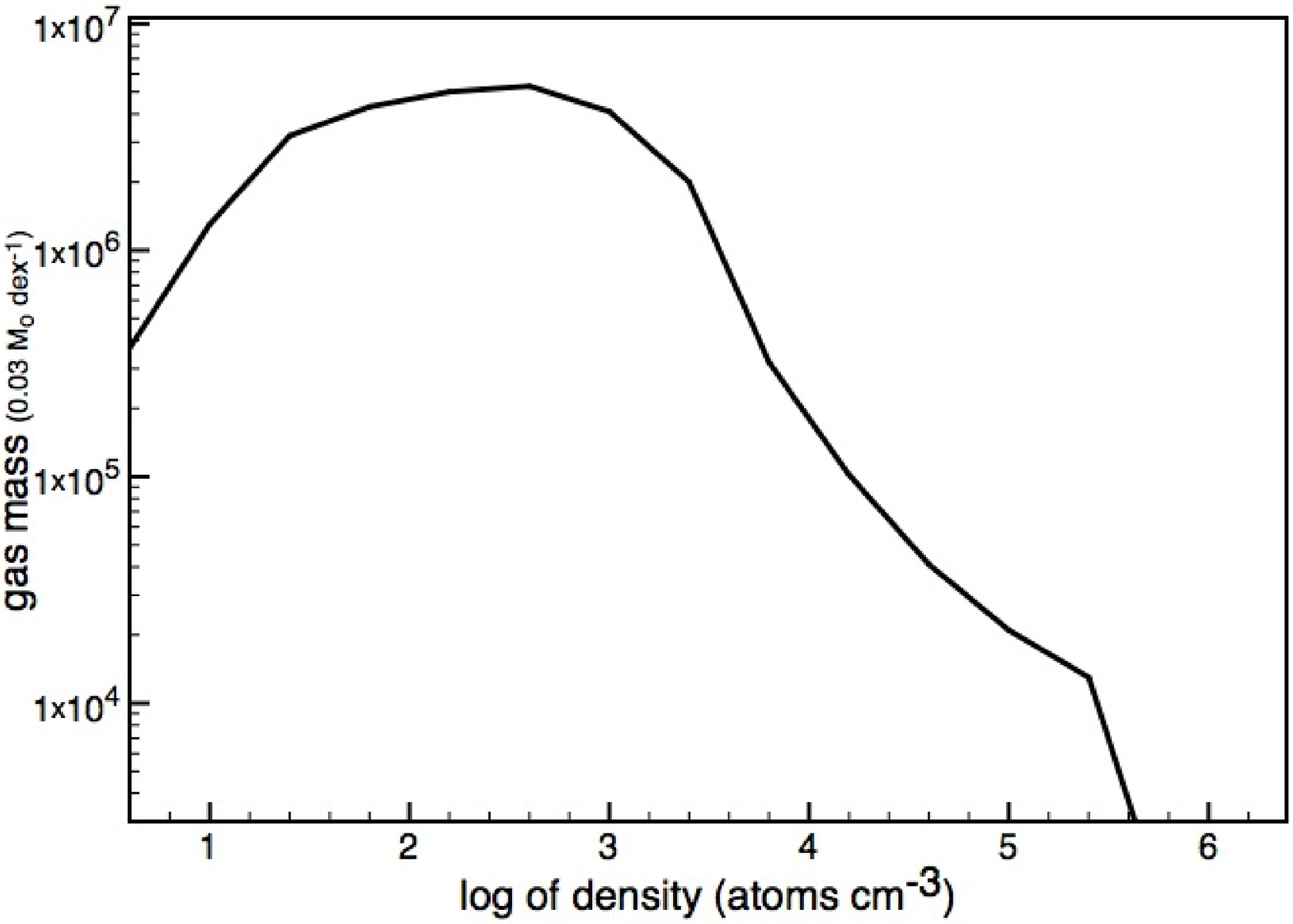} 
\caption{Zoom on a GMC in a simulation of a whole galaxy including gas, stars and dark matter from Bournaud et al. (2010a). The left panel shows the gas density over a portion of $300 \times 300$~pc in the face-on disk. The whole disk has a log-normal density PDF, the right panel shows the PDF for the single GMC shown to the left: it has a log-normal shape with a significant high-density tail. The high-density tail is where star formation takes place.}
\end{center}
\end{figure}

The process is almost entirely driven by turbulence and gravity, but  feedback is essential to maintain a steady state distribution. Without feedback, all gas mass will eventually end-up at the smallest scales of the turbulence cascade, in superdense bullets at the resolution limit. Feedback regulates the process by disrupting dense structures on the smallest scales and refuelling a turbulent steady state through larger-scale instabilities. This does not seem strongly affected by the hypothesis done on feedback, and adding HII region and radiative pressure feedback to supernovae feedback models could somewhat decrease the lifetime of gas clouds, but would not change the main properties of the ISM density distribution that are primarily controlled by the gravity-driven turbulence cascade. Although the small-scale substructures inside GMCs are probably still far from realistic (magnetic effects should eventually become important), it seems that the main GMC properties can be correctly modeled in simulations of whole galaxies or cosmological simulations, provided that:

-- The spatial resolution reaches about 1~pc: the gas disk scale height must be accurately resolved to capture a 3D turbulence cascade in the ISM, because there is no such cascade in 2D. A few resolution elements per scale height are not enough. One must resolve several Jeans length per disk scale height, so that instabilities can arise over-densities above and below the mid-plane and drive a vertical component of turbulent motions. As the thermal Jeans length itself is described by at least a few resolution elements to prevent artificial fragmentation, and the total, thermal+turbulent Jeans length is substantially larger -- at least 10-20 resolution elements for supersonic turbulence. This implies that the gas disk scale height has to be resolved with, say, 100-200 resolution elements:  the resolution for gravity {\em and} hydrodynamics for a typical $z=0$ galaxy needs to be around 1~pc. 

-- Star formation and energy feedback of some sort is included, so as to maintain a steady turbulent state where clouds do not all collapse down to the resolution limit.

-- Long-range gravitational forcings from old stellar populations are included (spiral density waves, bars, etc). Large-scale structures do participate to the turbulence cascade down to small scales, and the contribution of all stars to the disk stability in a combined $Q_{\mathrm gas+stars}$ parameter is about as important as that from the gas itself.

These requirements could in theory be achieved with any code, but in practice are probably easier to reach with oct-tree AMR codes such as RAMSES \citep{teyssier02}: grid refinements over complex systems such as GMCs in a spiral disk are optimized compared to patch-based and multi-grid techniques.

\section{Conclusions}

While the assembly of baryons into galaxies seems relatively well understood on large scales, the properties of galaxies in the simulations are still far from matching those of observed galaxy populations. There is increasing evidence that 
small-scale processes such as ISM turbulence, fragmentation, clustered star formation, are crucial to understand galaxy formation, in particular in the primordial high-redshift phases. They can have major effects on the global properties of each galaxy, and even on the global cosmological evolution of baryons. For instance, temptative explanations of the {\em downsizing} in the $\lambda$-CDM paradigm rely on AGN feedback \citep{scannapieco, dimatteo}. However, AGN fueling involves on a number processes inside the central kpc of galaxies that are completely unresolved in cosmological models (nuclear bars, resonnances, nuclear starbursts, mass loss from clusters, etc: see \citealt{combes}). 

Recent results and objective criteria to capture and correctly resolve the main star-forming regions of galaxies have been presented. They are unfortunately not reached in present-day cosmological simulations, especially at redshift zero, but recent improvements have been made in accounting for the multiphase ISM in galaxy formation models.

\end{document}